\documentclass[
 reprint,
 superscriptaddress,
 amsmath,amssymb,
 aps,
pra,
showkeys
]{revtex4-2}

\usepackage{soul}
\usepackage{float}
\usepackage{multibib}
\usepackage{xcolor}
\PassOptionsToPackage{hyphens}{url}
\usepackage{xurl} 
\usepackage{hyperref}
\hypersetup{
    colorlinks=true,
    breaklinks=true,
    linkcolor={blue!50!blue},
    citecolor={blue!50!blue},
    urlcolor={blue!80!black}
}

\usepackage{siunitx}
\usepackage{import}
\usepackage{placeins}
\usepackage{xcolor}
\usepackage{lmodern}
\usepackage[braket, qm]{qcircuit}
\usepackage{graphicx}
\usepackage{dcolumn}
\usepackage{bm}
\usepackage{comment}
\usepackage[english]{babel}
\usepackage{blindtext}
\usepackage{physics}  
\usepackage{multirow}
\usepackage{array}

\begin{document}

\title{QuaNTUM: A Modular Quantum Communication Testbed for Scalable Fiber and Satellite Integration}  

\newcommand{\tum}{Department of Computer Engineering, TUM School of Computation, Information and Technology, Technical University of Munich, 80333 Munich, Germany}
\newcommand{\mcqst}{Munich Center for Quantum Science and Technology (MCQST), 80799 Munich, Germany}
\newcommand{\jena}{Friedrich-Schiller-Universität Jena, Albert-Einstein-Straße 5, 07745 Jena, Germany}

\author{Julien Chénedé}
\email{julien.chenede@tum.de}
\affiliation{\tum}
\affiliation{\mcqst}

\author{Tjorben Matthes}
\affiliation{\tum}
\affiliation{\mcqst}

\author{Josefine Krause}
\affiliation{\tum}
\affiliation{\mcqst}
\affiliation{\jena}

\author{Asli Cakan}
\affiliation{\tum}
\affiliation{\mcqst}

\author{Tobias Vogl}
\email{tobias.vogl@tum.de}
\affiliation{\tum}
\affiliation{\mcqst}

\begin{abstract}
Secure communication is a cornerstone of modern society, underpinning everything from financial transactions to critical infrastructure. As classical encryption faces growing threats from advancing computational power, quantum communication offers a fundamentally secure alternative, leveraging the laws of physics to protect data.

We introduce QuaNTUM (Quantum Network at the Technical University of Munich), a modular and extensible quantum communication testbed designed to enable scalable, flexible, and secure quantum communication across fiber-based campus networks and satellite-ground links. QuaNTUM integrates early deployments of solid-state quantum emitters in small satellites, bridging terrestrial and free-space channels. As an open-access platform, it supports experimental quantum communication protocols, quantum device benchmarking, and hybrid network integration.

The terrestrial network connects universities and research institutes across the high-tech campus in Garching near Munich, using single-mode fibers in a star-shaped topology. Each node features polarization-maintaining components, multiplexers, and time-synchronized analysis modules, with a central switching hub for dynamic routing. Active polarization control and real-time feedback ensure low error rates and stable qubit transmission, enabling high-fidelity quantum key distribution (QKD) and entanglement distribution.

A central focus of QuaNTUM is its use of deterministic single-photon sources based on optically active defects, such as in hexagonal boron nitride (hBN) or excited erbium atoms. These sources are currently being deployed in orbit, marking one of the first demonstrations of a solid-state quantum emitter in space. By uniting fiber and free-space links with scalable hardware and open protocols, QuaNTUM offers a basis that could adapt to future hybrid quantum networks, supporting both current testbed research and the long-term vision of a global quantum internet.
\end{abstract}


\maketitle

\section{Introduction}

The arrival of the quantum computer threatens today’s encrypted networks by efficiently solving complex mathematical problems (like factoring large integers), on which nowadays most of our secured communication system are relying on \cite{Gisin}. One replacement candidate is quantum communication which guarantees a safe communication method against any eavesdropper or even the operator of the network. Several projects were done during the past years proving the feasibility of such links using long distance communication via satellites \cite{micius} and shorter ones via fibers \cite{Qinternet}. Before quantum repeaters become technically feasible and enable a full "quantum internet", hybrid networks can be created using both fiber and satellite links, with satellites serving as trusted nodes. Ultimately, QuaNTUM aims to enable long-distance quantum networking, linking remote quantum processors and supporting reliable quantum information exchange.

Quantum key distribution (QKD) is a remarkable technology: it uses the fundamental laws of physics to secure communication. By encoding keys onto single photons, it makes eavesdropping impossible, thanks to the no-cloning theorem \cite{Gisin} and the fact that any attempt to measure the photons disturbs them. Yet, a limitation remains: when we rely only on fiber optics, QKD hits a wall after a few hundred kilometers, or even less for some implementations. The signal weakens exponentially, and unlike classical signals, amplification alone is impossible.

Realizing a true "quantum internet" relies on quantum repeaters, which are still nowadays at an early stage \cite{Qinternet}. In the meantime, there is a workaround: a hybrid approach that combines the best of both worlds. Satellites can bridge vast distances through free space, where signal loss is almost negligible above 10 kilometers, while fiber networks handle the last kilometers, connecting buildings, campuses, or even cities.

Despite these advancements, a critical gap remains: the absence of a large-scale, real-world open testbed that integrates these technologies into a cohesive framework.
The QuaNTUM project addresses this integration challenge by establishing a star topology fiber network across the Garching campus near Munich. This infrastructure interconnects various institutes and laboratories, enabling shared access to quantum resources. Unlike trusted-node QKD networks that have already been demonstrated in Boston, Vienna, Tokyo, and China, QuaNTUM is designed as an open research testbed that directly connects laboratories for end-to-end quantum experiments without relying on intermediate trusted relays. Researchers can thus conduct experiments involving single-photon emission, for instance, from hBN \cite{Tobias} color centers or erbium-doped materials \cite{Andreas}, optical quantum memories \cite{Memory,Asli}, and diverse QKD protocols \cite{Qinternet}, as well as investigate quantum entanglement distribution \cite{Neumann2022}.

In a subsequent phase, the network will be extended to Munich’s city center and to a ground station to support optical satellite links. Preparatory work is already underway with the low-Earth-orbit CubeSat QUICK³ \cite{QUICK3}, which is performing foundational quantum optics experiments in microgravity.
Upon completion of the commissioning phase, the QuaNTUM network will be made accessible to the Munich Center for Quantum Science and Technology (MCQST), the Munich Quantum Valley (MQV), and external partners. This will facilitate collaborative research in areas such as secure quantum communication, quantum sensing, and the interfacing of quantum memories.

\section{QuaNTUM Testbed Architecture}

\subsection{Fiber‐Optic Backbone}

\begin{figure*}[ht]
    \includegraphics[width=\textwidth]{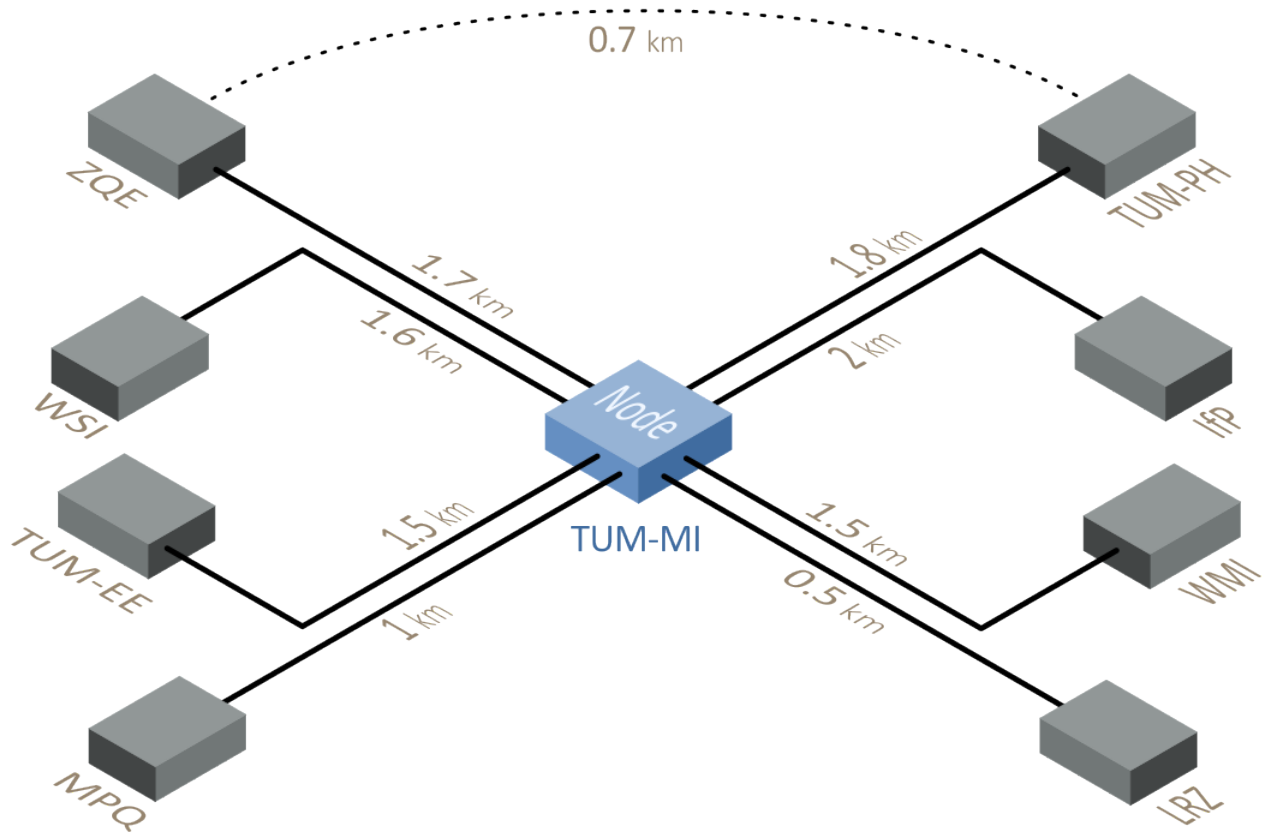}
        \caption{Topology of the QuaNTUM testbed on the Garching campus near Munich: star‐shaped fiber connections using SMF-28 ultra fibers, 780-HP and 1060-XP between the department of physics (TUM‐PH), Electrical Engineering (TUM‐EE), the Max Planck Institute of Quantum Optics (MPQ), the Leibniz Supercomputing Centre (LRZ), the Walter Schottky Institut (WSI), the Walther-Meißner-Institut (WMI) and with the central node in School of Computation, Information and Technology (TUM-MI). An additional special cable with only SMF-28 ultra fibers is connecting directly ZQE and TUM-PH (dotted-line).
         -- For geographic reference, see the map: \href{https://v.bayern.de/N74hf}{https://v.bayern.de/N74hf}
         -- ©Bayerische Vermessungsverwaltung 2025
        }
     \label{network}
\end{figure*}

The QuaNTUM fiber-optic backbone is implemented as a star topology single-mode network centered at the TUM-MI node (Departments of Mathematics and Computer Science), roughly located in the center of the campus. Some standard Telecom-Band SMF-28 fibers form the primary campus backbone and interconnect the principal sites described in \textbf{Fig. \ref{network}} with typical link lengths of 1\,km to 2\,km. At the central node a quantum reconfigurable add–drop multiplexer (q-ROADM) is specified to provide wavelength de/multiplexing, low-loss optical switching and a consolidated point for polarization control hardware and timing distribution. The network design preserves the option of direct fiber-to-fiber links to minimize insertion loss for critical quantum channels.

Cable composition and connectorization are chosen to balance compatibility with telecom infrastructure and the specific wavelength requirements of quantum systems. Each main cable contains multiple SMF-28 fibers (optimized for telecom O-/C-bands at 1330 nm and 1550 nm) together with dedicated short-wavelength fibers (780-HP and 1060-XP) to support key quantum technologies \cite{nos}:

\begin{itemize}
    \item Solid-state qubits: Color centers in diamond (e.g. SiV centers at ~738\,nm \cite{Bradac2019}) and silicon carbide (e.g. silicon vacancy centers emitting at longer near-infrared wavelengths \cite{Castelletto_2021}) require near-visible to near-infrared wavelengths, but fiber attenuation below 780\,nm is prohibitive for kilometer-scale links—hence the 780\,nm lower bound in our design.
    \item Quantum memories: Rare-earth-doped crystals (e.g. Tm$^{3+}$:Y$_2$SiO$_5$ at 793\,nm \cite{Simon2010}) and alkali vapors (e.g. Rb D2 transitions at 780\,nm \cite{nos}) often operate in the 700–800\,nm range, motivating the inclusion of 780-HP fibers.
    \item Telecom compatibility: Long-distance quantum communication leverages the near-infrared (NIR) free-space window and telecom O-/C-bands (1330/1550\,nm).
\end{itemize}

E2000 APC simplex terminations are specified for quantum ports to minimize back-reflection and allow individual fiber access. Loss budgeting and noise mitigation are central design constraints. The end-to-end optical loss model includes fiber attenuation, splice loss, connector mated loss, and insertion loss of passive/active elements (Wavelength Division Multiplexing \cite{WDM}, switches, polarizers).
To protect single-photon channels from classical noise, high-power classical signals are separated spectrally or routed on separate fibers; APC connectors are used to reduce back reflections. For polarization-encoded protocols the backbone incorporates motorized fiber polarization controllers (FPCs) and software feedback to maintain state fidelity under environmental perturbations. Market-ready devices devices can provide polarization compensation such as silicon-based decoder \cite{siliconbased}, Pockels cell \cite{Caputo2022} and fiber squeezing devices \cite{squeezing}.

\subsection{Timing and Control}
Synchronization and deterministic control are fundamental to all QuaNTUM experiments: they enable reliable coincidence detection, accurate protocol timing, and automated feedback for polarization and routing. The timing architecture therefore combines a dedicated classical reference channel, local high-resolution time-tagging at each measurement node, and a coordinated control plane that drives active hardware and manages experiment scheduling.

A low-jitter classical reference channel is provisioned alongside the quantum wavelengths to distribute a 10\,MHz clock and a 1 pulse per second (PPS) marker. This provides a stable, common time base and avoids many limitations of satellite-based references for short-range campus networks. Each polarization analysis module (PAM) is paired with a synchronized time-tagger capable of sub-100-ps resolution for arrival-time stamping; these local time stamps are essential for coincidence analysis, quantum bit error ratio (QBER) estimation and post-processing.

For deterministic, low-latency distribution and scalable network timing, the testbed will evaluate precision time protocols such as White Rabbit \cite{whiterabbit} because they provide sub-nanosecond synchronization and known latency bounds over fiber links.

The control plane is implemented as a hybrid of FPGA/embedded control at the hardware layer and a centralized scheduler at the management layer. FPGA units or microcontrollers co-located with the q-ROADM and at user nodes perform deterministic control of optical switches, electro-optic modulators and motorized FPCs with microsecond actuation times. Timing-critical loops, for example, live polarization compensation, run locally at each node (fast perturbation, local evaluation, accept/reject), while higher-level policies (when to perform a full reference recalibration, scheduling of satellite passes) are managed centrally.

Practical implementation requires explicit calibration and monitoring: per-link fixed delays must be measured and recorded (optical time-domain reflectometer and two-way time transfer), time-tagger offsets and cable dispersive effects characterized, and ongoing telemetry (detector counts, QBER, temperature) integrated into the control dashboard. These elements together ensure that QuaNTUM achieves the temporal precision and operational automation necessary for reliable QKD and entanglement distribution experiments across a campus-scale hybrid fiber network.

\subsection{Entanglement and QKD Layers}
From a security standpoint, protocols divide into device-independent (theoretically secure against all attacks, with no hardware assumptions) and device-dependent (relying on trusted measurement tools but compatible with off-the-shelf components). In practical terms, implementations split between discrete-variable (e.g. polarized photons) and continuous-variable (e.g. coherent states) approaches \cite{QKD}. QKD protocols fall into three primary types (Table 1)\cite{QKD}:

\begin{table*}[ht]
    \centering
    \caption{QKD protocol classification by structure}
    \begin{tabular*}{\linewidth}{@{\extracolsep{\fill}} c c @{}}
        \hline
        \textbf{Type} & \textbf{Key Mechanism} \\
        \hline
        Prepare-and-Measure (PM) &
        Alice prepares quantum states; Bob measures them to generate a shared key. \\

        Entanglement-Based &
        Alice and Bob exploit entangled particle correlations for key distribution. \\

        Measurement-Device-Independent &
        Security remains intact even if measurement devices are untrusted. \\ \hline

    \end{tabular*}
    
    \label{tab:qkd_protocols}
\end{table*}

 QuaNTUM is implementing two complementary quantum channel layers on the same physical backbone: a wavelength-multiplexed entanglement distribution layer for multi-user and a PM-QKD layer that supports standard schemes.

For entanglement distribution, a broadband spontaneous-parametric-down-conversion (SPDC) source in a Sagnac configuration \cite{Sagnac} is specified as the primary photon-pair generator. The SPDC output is spectrally partitioned by a demultiplexer; the q-ROADM in the central node assigns channel pairs to endpoint links, producing a reconfigurable, wavelength-multiplexed entanglement fabric. At each user endpoint, polarization analysis modules (PAMs) implement random-basis selection and perform polarization projection and detection. Time-stamping of events with sub-100-ps resolution enables coincidence identification and correlation analysis for entanglement verification. The architecture is designed to support multi-hop protocols \cite{whiterabbit} by routing idler photons to intermediate Bell-state measurement stations when required, enabling tests of networked entanglement distribution.

The PM layer will support different standard QKD implementations (for example B92, BB84, SARG04 \cite{QKD}).
Deterministic single-photon emitters based on localized defects in hBN (at 575\,nm) are primarily intended for short-reach or on-chip experiments, as their visible-band emission is incompatible with long-distance transmission over telecom-band fibers. For integration with the testbed’s backbone, a frequency conversion strategy is employed: Visible photons are converted to telecom bands via nonlinear optics \cite{FC} before entering the network.

For long-distance operation, the testbed provisions alternative sources (e.g. telecom-compatible Spontaneous Parametric Down-Conversion pairs \cite{SPDC} or InGaAs-based single-photon emitters\cite{InGaAs}). Real-time operational control closes the loop for secure operation: measured QBER and coincidence statistics drive local feedback (motorized fiber polarization controllers and software routines) for live polarization compensation and calibration.

\section{Single‐Photon Emitters in hBN}

\subsection{Deterministic Fabrication}

In QuaNTUM we implement deterministic fabrication by localized electron-beam irradiation of multilayer hBN using a standard scanning electron microscope (SEM) \cite{Anand}. Exposures are targeted to sub-micron coordinates to create emitter arrays with high lateral precision; process parameters (beam energy, current, dose and spot dwell time) are optimized to maximize yield and spectral reproducibility. The produced emitters show a reproducible zero-phonon line (ZPL) near 575\,nm and are consistent with carbon-related defect complexes identified by density-functional calculations\cite{Cholsuk2023}. This approach is compatible with wafer-scale patterning workflows and with subsequent on-chip or micro-optical integration because the emitter position is known \textit{a priori}. Scalability hinges on automation of exposure patterning, batch handling of hBN flakes or films, and yield optimization to reduce per-device screening. For production of fiber-coupled modules or on-chip arrays, deterministic sites are pre-aligned to lithographically defined photonic structures (waveguides, cavities, grating couplers) so that subsequent assembly and coupling are mainly mechanical rather than optical alignment tasks.

\subsection{Optical Performance}

\begin{figure*}[ht]
  \includegraphics[width=\textwidth]{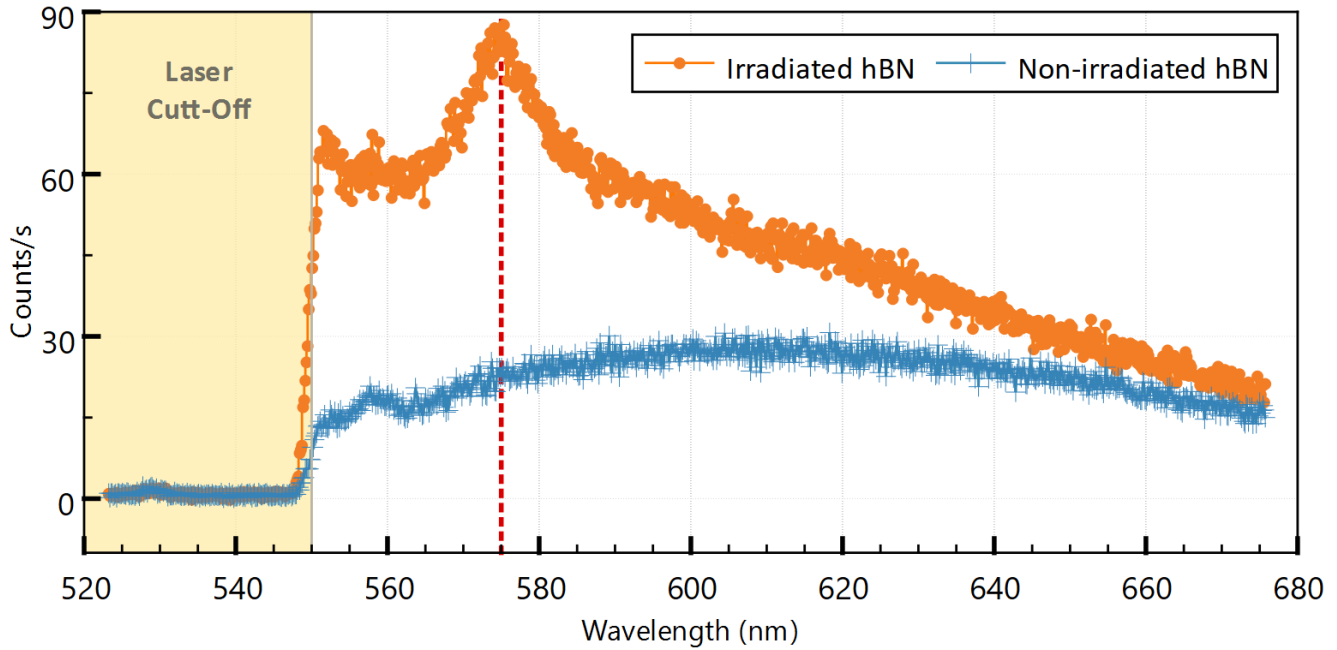}
  \caption{Room-temperature photoluminescence spectrum of a single hBN emitter fabricated by localized electron irradiation, featuring peak emission near 575\,nm (orange line). Comparison with a non-irradiated hBN (blue line). Laser excitation at 530\,nm and long-pass filter at 550\,nm.}
  \label{hBN}
\end{figure*}

hBN hosts a diverse family of quantum emitters with zero-phonon lines (ZPLs) spanning UV to NIR wavelengths \cite{databaseNos}, enabling compatibility with a wide range of quantum systems. For this work, we focus on defects produced by localized irradiation that, under 530\,nm femtosecond laser excitation, exhibit a peak emission at 575\,nm with a typical linewidth (FWHM) on the order of tens of nanometers at room temperature (\textbf{Fig. \ref{hBN}}). The second-order correlation function $g^{(2)}(0)<0.1$ confirms single-photon emission \cite{Anand}, while the reported excited-state lifetime of 3.83(1)\,ns supports high repetition-rate operation \cite{Anand}. Long-term photostability has been observed in many sites even 3 years post-irradiation, though device-level qualification—including performance under variable temperature and optical pump conditions—remains necessary for network deployment or space applications. Visible-band ZPLs and room-temperature linewidths are advantageous for bright, high-rate operation but pose compatibility challenges with standard SMF-28 telecom infrastructure and with narrowband quantum memories. To address these issues, two engineering strategies are relevant: (1) spectral narrowing and tuning via photonic engineering (microcavities, Purcell enhancement, strain or electric-field tuning) to reduce the effective linewidth and stabilize the ZPL; and (2) quantum frequency conversion (QFC) to translate emitted photons to telecom wavelengths for low-loss fiber transmission or to memory-matched atomic transitions. Microcavity coupling can also increase the collection efficiency and reduce the required source brightness for a target key rate or coincidence rate, while QFC preserves quantum statistics if conversion noise is sufficiently suppressed.

\subsection{Integrated photonics}
Integration of hBN emitters with photonic platforms (e.g. fiber integrated platforms \cite{Vogl_2017}) addresses three objectives: improve collection/coupling efficiency into single-mode fibers, control emission dynamics (Purcell effect) and enable on-chip routing and switching for compact, stable modules \cite{Tobias}. In practice, deterministic emitter sites are aligned to on-chip waveguides, microcavities or grating couplers using lithographic markers and high-precision assembly. Finite-difference time-domain method (FDTD) simulations guide the resonator and waveguide design to maximize coupling efficiency and to position cavity resonances relative to the emitter ZPL \cite{Finley}. Thanks to photonic integration, stricter constraints are typically designed for satellite deployment. The QUICK³ payload strategy uses a miniaturized version of a different emitter platform based on MoSe$_2$ \cite{QUICK3}: the chip is mounted within a compact, thermally managed optical bench that includes the pump laser, stabilization optics, fiber-coupling stage and pointing interface. Packaging choices emphasize mechanical resilience (vibration and shock tolerance), thermal control over the expected orbital temperature range, radiation tolerance of active components, and minimal moving parts.

\section{Conclusion}
In conclusion, the QuaNTUM project addresses a critical gap in the field of quantum communication by providing a large-scale, real-world open testbed that integrates fiber and satellite links into a cohesive framework. This platform, akin to a campus-wide hybrid network, offers researchers and engineers a controlled yet dynamic environment to experimentally validate, refine, and benchmark system performance under realistic conditions. By leveraging advanced technologies such as solid-state quantum emitters and polarization control, QuaNTUM aims to overcome the limitations of current quantum communication systems. The successful integration of these technologies could pave the way for a global quantum internet, enabling secure communication and advanced quantum applications. Future work will focus on expanding the network, improving the performance of quantum emitters, and exploring new applications in quantum communication and sensing.

\section*{Data availability}
All data from this work is available from the authors upon reasonable request.

\section*{Notes}
The authors declare that they have no competing interests relevant to the content of this article. This is the author’s accepted manuscript of a paper published in Communications in Computer and Information Science (CCIS), vol. 2743, Springer (2026).

\begin{acknowledgments}
This research is part of the Munich Quantum Valley, which is supported by the Bavarian state government with funds from the Hightech Agenda Bayern Plus. This work was funded by the Deutsche Forschungsgemeinschaft (DFG, German Research Foundation) under Germany’s Excellence Strategy- EXC-2111-390814868 (MCQST). The authors acknowledge support from the Federal Ministry of Research, Technology and Space (BMFTR) under grant number 13N16292 (ATOMIQS).
\end{acknowledgments}


\bibliography{main}

@article{Gisin,
  title = {Quantum cryptography},
  author = {Gisin, Nicolas and Ribordy, Gr\'egoire and Tittel, Wolfgang and Zbinden, Hugo},
  journal = {Rev. Mod. Phys.},
  volume = {74},
  issue = {1},
  pages = {145--195},
  numpages = {0},
  year = {2002},
  month = {Mar},
  publisher = {American Physical Society},
  doi = {10.1103/RevModPhys.74.145},
  url = {https://link.aps.org/doi/10.1103/RevModPhys.74.145}
}

@article{micius,
  title = {Micius quantum experiments in space},
  author = {Lu, Chao-Yang and Cao, Yuan and Peng, Cheng-Zhi and Pan, Jian-Wei},
  journal = {Rev. Mod. Phys.},
  volume = {94},
  issue = {3},
  pages = {035001},
  numpages = {46},
  year = {2022},
  month = {Jul},
  publisher = {American Physical Society},
  doi = {10.1103/RevModPhys.94.035001},
  url = {https://link.aps.org/doi/10.1103/RevModPhys.94.035001}
}

@article{Qinternet,
  author={Granados, German and Velasquez, Washington and Cajo, Ricardo and Antonieta-Alvarez, Maria},
  journal={IEEE Access}, 
  title={Quantum Key Distribution in Multiple Fiber Networks and Its Application in Urban Communications: A Comprehensive Review}, 
  year={2025},
  volume={13},
  number={},
  pages={100446-100461},
  doi={10.1109/ACCESS.2025.3577086}
}

@article{Tobias,
author = {Vogl, Tobias and Lecamwasam, Ruvi and Buchler, Ben C. and Lu, Yuerui and Lam, Ping Koy},
title = {Compact Cavity-Enhanced Single-Photon Generation with Hexagonal Boron Nitride},
journal = {ACS Photonics},
volume = {6},
number = {8},
pages = {1955-1962},
year = {2019},
doi = {10.1021/acsphotonics.9b00314},
url = { 
        https://doi.org/10.1021/acsphotonics.9b00314
},
eprint = { 
    
        https://doi.org/10.1021/acsphotonics.9b00314
}
}

@article{Andreas,
  title = {Narrow Optical Transitions in Erbium-Implanted Silicon Waveguides},
  author = {Gritsch, Andreas and Weiss, Lorenz and Fr\"uh, Johannes and Rinner, Stephan and Reiserer, Andreas},
  journal = {Phys. Rev. X},
  volume = {12},
  issue = {4},
  pages = {041009},
  numpages = {19},
  year = {2022},
  month = {Oct},
  publisher = {American Physical Society},
  doi = {10.1103/PhysRevX.12.041009},
  url = {https://link.aps.org/doi/10.1103/PhysRevX.12.041009}
}

@article{Memory,
    author={Lvovsky, Alexander I.
    and Sanders, Barry C.
    and Tittel, Wolfgang},
    title={Optical quantum memory},
    journal={Nature Photonics},
    year={2009},
    month={Dec},
    day={01},
    volume={3},
    number={12},
    pages={706-714},
    issn={1749-4893},
    doi={10.1038/nphoton.2009.231},
    url={https://doi.org/10.1038/nphoton.2009.231}
}

@article{Asli,
    author = {Çakan, Aslı and Cholsuk, Chanaprom and Gale, Angus and Kianinia, Mehran and Paçal, Serkan and Ateş, Serkan and Aharonovich, Igor and Toth, Milos and Vogl, Tobias},
    title = {Quantum Optics Applications of Hexagonal Boron Nitride Defects},
    journal = {Advanced Optical Materials},
    volume = {13},
    number = {7},
    pages = {2402508},
    doi = {https://doi.org/10.1002/adom.202402508},
    year = {2025}
}

@Article{nos,
    AUTHOR = {Cholsuk, Chanaprom and Suwanna, Sujin and Vogl, Tobias},
    TITLE = {Tailoring the Emission Wavelength of Color Centers in Hexagonal Boron Nitride for Quantum Applications},
    JOURNAL = {Nanomaterials},
    VOLUME = {12},
    YEAR = {2022},
    NUMBER = {14},
    ARTICLE-NUMBER = {2427},
    url = {https://www.mdpi.com/2079-4991/12/14/2427},
    PubMedID = {35889651},
    ISSN = {2079-4991},
    DOI = {10.3390/nano12142427}
}

@article{QUICK3,
    author = {Ahmadi, Najme and Schwertfeger, Sven and Werner, Philipp and Wiese, Lukas and Lester, Joseph and Da Ros, Elisa and Krause, Josefine and Ritter, Sebastian and Abasifard, Mostafa and Cholsuk, Chanaprom and Krämer, Ria G. and Atzeni, Simone and Gündoğan, Mustafa and Sachidananda, Subash and Pardo, Daniel and Nolte, Stefan and Lohrmann, Alexander and Ling, Alexander and Bartholomäus, Julian and Corrielli, Giacomo and Krutzik, Markus and Vogl, Tobias},
    title = {QUICK³ - Design of a Satellite-Based Quantum Light Source for Quantum Communication and Extended Physical Theory Tests in Space},
    journal = {Advanced Quantum Technologies},
    volume = {7},
    number = {4},
    pages = {2300343},
    doi = {https://doi.org/10.1002/qute.202300343},
    year = {2024}
}

@article{Neumann2022,
    author={Neumann, Sebastian Philipp
    and Buchner, Alexander
    and Bulla, Lukas
    and Bohmann, Martin
    and Ursin, Rupert},
    title={Continuous entanglement distribution over a transnational 248{\thinspace}km fiber link},
    journal={Nature Communications},
    year={2022},
    month={Oct},
    day={17},
    volume={13},
    number={1},
    pages={6134},
    issn={2041-1723},
    doi={10.1038/s41467-022-33919-0},
    url={https://doi.org/10.1038/s41467-022-33919-0}
}

@article{WDM,
  title = {Quantum wavelength-division-multiplexing and multiple-access communication systems and networks: Global and unified approach},
  author = {Bathaee, Marzieh and Rezai, Mohammad and Salehi, Jawad A.},
  journal = {Phys. Rev. A},
  volume = {107},
  issue = {1},
  pages = {012613},
  numpages = {17},
  year = {2023},
  month = {Jan},
  publisher = {American Physical Society},
  doi = {10.1103/PhysRevA.107.012613},
  url = {https://link.aps.org/doi/10.1103/PhysRevA.107.012613}
}

@article{siliconbased,
    title = {Silicon-based decoder for polarization-encoding quantum key distribution},
    journal = {Chip},
    volume = {2},
    number = {1},
    pages = {100039},
    year = {2023},
    issn = {2709-4723},
    doi = {https://doi.org/10.1016/j.chip.2023.100039},
    url = {https://www.sciencedirect.com/science/article/pii/S2709472323000023},
    author = {Yongqiang Du and Xun Zhu and Xin Hua and Zhengeng Zhao and Xiao Hu and Yi Qian and Xi Xiao and Kejin Wei}
}

@article{Caputo2022,
    author={Caputo, Carlo
    and Simoni, Mario
    and Cirillo, Giovanni Amedeo
    and Turvani, Giovanna
    and Zamboni, Maurizio},
    title={A simulator of optical coherent-state evolution in quantum key distribution systems},
    journal={Optical and Quantum Electronics},
    year={2022},
    month={Sep},
    day={13},
    volume={54},
    number={11},
    pages={689},
    issn={1572-817X},
    doi={10.1007/s11082-022-04041-8},
    url={https://doi.org/10.1007/s11082-022-04041-8}
}

@article{squeezing,
    author = {F. Kaiser and B. Fedrici and A. Zavatta and V. D'Auria and S. Tanzilli},
    journal = {Optica},
    number = {4},
    pages = {362--365},
    publisher = {Optica Publishing Group},
    title = {A fully guided-wave squeezing experiment for fiber quantum networks},
    volume = {3},
    month = {Apr},
    year = {2016},
    url = {https://opg.optica.org/optica/abstract.cfm?URI=optica-3-4-362},
    doi = {10.1364/OPTICA.3.000362},
}

@article{whiterabbit,
  author={Alshowkan, Muneer and Lukens, Joseph M. and Lu, Hsuan-Hao and Peters, Nicholas A.},
  journal={Journal of Lightwave Technology}, 
  title={Resilient Entanglement Distribution in a Multihop Quantum Network}, 
  year={2025},
  volume={},
  number={},
  pages={1-8},
  doi={10.1109/JLT.2025.3581220}
}

@article{QKD,
    author={Kumar, Mandeep
    and Mondal, Bhaskar},
    title={A brief review on Quantum Key Distribution Protocols},
    journal={Multimedia Tools and Applications},
    year={2025},
    month={Aug},
    day={01},
    volume={84},
    number={27},
    pages={33267-33306},
    issn={1573-7721},
    doi={10.1007/s11042-024-20535-x}
}

@Article{SPDC,
    author={Ikuta, Rikizo
    and Kusaka, Yoshiaki
    and Kitano, Tsuyoshi
    and Kato, Hiroshi
    and Yamamoto, Takashi
    and Koashi, Masato
    and Imoto, Nobuyuki},
    title={Wide-band quantum interface for visible-to-telecommunication wavelength conversion},
    journal={Nature Communications},
    year={2011},
    month={Nov},
    day={15},
    volume={2},
    number={1},
    pages={537},
    issn={2041-1723},
    doi={10.1038/ncomms1544},
    url={https://doi.org/10.1038/ncomms1544}
}

@Article{InGaAs,
    author={Zhang, Jun
    and Itzler, Mark A.
    and Zbinden, Hugo
    and Pan, Jian-Wei},
    title={Advances in InGaAs/InP single-photon detector systems for quantum communication},
    journal={Light: Science {\&} Applications},
    year={2015},
    month={May},
    day={01},
    volume={4},
    number={5},
    pages={e286-e286},
    issn={2047-7538},
    doi={10.1038/lsa.2015.59},
    url={https://doi.org/10.1038/lsa.2015.59}
}

@Article{Bradac2019,
    author={Bradac, Carlo
    and Gao, Weibo
    and Forneris, Jacopo
    and Trusheim, Matthew E.
    and Aharonovich, Igor},
    title={Quantum nanophotonics with group IV defects in diamond},
    journal={Nature Communications},
    year={2019},
    month={Dec},
    day={09},
    volume={10},
    number={1},
    pages={5625},
    issn={2041-1723},
    doi={10.1038/s41467-019-13332-w},
    url={https://doi.org/10.1038/s41467-019-13332-w}
}

@article{Castelletto_2021,
    doi = {10.1088/2633-4356/abe04a},
    url = {https://dx.doi.org/10.1088/2633-4356/abe04a},
    year = {2021},
    month = {mar},
    publisher = {IOP Publishing},
    volume = {1},
    number = {2},
    pages = {023001},
    author = {Castelletto, Stefania},
    title = {Silicon carbide single-photon sources: challenges and prospects},
    journal = {Materials for Quantum Technology}
}

@Article{Simon2010,
    author={Simon, C.
    and Afzelius, M.
    and Appel, J.
    and Boyer de la Giroday, A.
    and Dewhurst, S. J.
    and Gisin, N.
    and Hu, C. Y.
    and Jelezko, F.
    and Kr{\"o}ll, S.
    and M{\"u}ller, J. H.
    and Nunn, J.
    and Polzik, E. S.
    and Rarity, J. G.
    and De Riedmatten, H.
    and Rosenfeld, W.
    and Shields, A. J.
    and Sk{\"o}ld, N.
    and Stevenson, R. M.
    and Thew, R.
    and Walmsley, I. A.
    and Weber, M. C.
    and Weinfurter, H.
    and Wrachtrup, J.
    and Young, R. J.},
    title={Quantum memories},
    journal={The European Physical Journal D},
    year={2010},
    month={May},
    day={01},
    volume={58},
    number={1},
    pages={1-22},
    issn={1434-6079},
    doi={10.1140/epjd/e2010-00103-y},
    url={https://doi.org/10.1140/epjd/e2010-00103-y}
}

@article{FC,
  title = {Single Photon Frequency Conversion for Frequency Multiplexed Quantum Networks in the Telecom Band},
  author = {Fisher, Paul and Cernansky, Robert and Haylock, Ben and Lobino, Mirko},
  journal = {Phys. Rev. Lett.},
  volume = {127},
  issue = {2},
  pages = {023602},
  numpages = {6},
  year = {2021},
  month = {Jul},
  publisher = {American Physical Society},
  doi = {10.1103/PhysRevLett.127.023602},
  url = {https://link.aps.org/doi/10.1103/PhysRevLett.127.023602}
}

@article{databaseNos,
    author = {Cholsuk, Chanaprom and Zand, Ashkan and {\c{C}}akan, Aslı and Vogl, Tobias},
    title = {The hBN Defects Database: A Theoretical Compilation of Color Centers in Hexagonal Boron Nitride},
    journal = {The Journal of Physical Chemistry C},
    volume = {128},
    number = {30},
    pages = {12716-12725},
    year = {2024},
    doi = {10.1021/acs.jpcc.4c03404},
    url = {
            https://doi.org/10.1021/acs.jpcc.4c03404
    },
    eprint = { 
            https://doi.org/10.1021/acs.jpcc.4c03404
    }
}

@article{Vogl_2017,
    doi = {10.1088/1361-6463/aa7839},
    url = {https://dx.doi.org/10.1088/1361-6463/aa7839},
    year = {2017},
    month = {Jun},
    publisher = {IOP Publishing},
    volume = {50},
    number = {29},
    pages = {295101},
    author = {Vogl, Tobias and Lu, Yuerui and Koy Lam, Ping},
    title = {Room temperature single photon source using fiber-integrated hexagonal boron nitride},
    journal = {Journal of Physics D: Applied Physics}
}

@article{Sagnac,
  title = {Demonstration of a Quantum Switch in a Sagnac Configuration},
  author = {Str\"omberg, Teodor and Schiansky, Peter and Peterson, Robert W. and Quintino, Marco T\'ulio and Walther, Philip},
  journal = {Phys. Rev. Lett.},
  volume = {131},
  issue = {6},
  pages = {060803},
  numpages = {6},
  year = {2023},
  month = {Aug},
  publisher = {American Physical Society},
  doi = {10.1103/PhysRevLett.131.060803},
  url = {https://link.aps.org/doi/10.1103/PhysRevLett.131.060803}
}

@article{Anand,
    author = {Kumar, Anand and Cholsuk, Chanaprom and Zand, Ashkan and Mishuk, Mohammad N. and Matthes, Tjorben and Eilenberger, Falk and Suwanna, Sujin and Vogl, Tobias},
    title = {Localized creation of yellow single photon emitting carbon complexes in hexagonal boron nitride},
    journal = {APL Materials},
    volume = {11},
    number = {7},
    pages = {071108},
    year = {2023},
    month = {Jul},
    issn = {2166-532X},
    doi = {10.1063/5.0147560},
    url = {https://doi.org/10.1063/5.0147560}
}

@article{Finley,
    url = {https://doi.org/10.1515/nanoph-2023-0347},
    title = {Thickness insensitive nanocavities for 2D heterostructures using photonic molecules},
    author = {Peirui Ji and Chenjiang Qian and Jonathan J. Finley and Shuming Yang},
    pages = {3501--3510},
    volume = {12},
    number = {17},
    journal = {Nanophotonics},
    doi = {doi:10.1515/nanoph-2023-0347},
    year = {2023},
    lastchecked = {2025-09-14}
}

@Article{Cholsuk2023,
author={Cholsuk, Chanaprom and Suwanna, Sujin and Vogl, Tobias},
title={Comprehensive Scheme for Identifying Defects in Solid-State Quantum Systems},
journal={The Journal of Physical Chemistry Letters},
year={2023},
month={Jul},
day={27},
publisher={American Chemical Society},
volume={14},
number={29},
pages={6564-6571},
doi={10.1021/acs.jpclett.3c01475},
url={https://doi.org/10.1021/acs.jpclett.3c01475}
}

\clearpage

\end{document}